\documentclass[useAMS,usenatbib]{mn2e}
\usepackage{times}
\usepackage{url}
\usepackage{graphicx}
\graphicspath{ {figs/}}

\newcommand{\ro}{Rev.~1491--1494}
\newcommand{\rn}{Rev.~1971--1974}
\newcommand\ion[2]{#1\,{\scshape{#2}}}

\title[\textsl{XMM-Newton} Observation of 1H0707$-$495]{Spectral
  Analysis of 1H0707$-$495 with \textsl{XMM-Newton}} \author[T. Dauser
et al.]{T.\ Dauser$^{1}$\thanks{E-mail:
    thomas.dauser@sternwarte.uni-erlangen.de}, J. Svoboda$^{2,3}$, N.
  Schartel$^{3}$, J. Wilms$^{1}$, M. Dov\v{c}iak$^{2}$, M. Ehle$^{3}$,
  \newauthor V. Karas$^{2}$,
  M. Santos-Lle\'{o}$^{3}$, and H. L. Marshall$^{4}$ \\
  $^{1}$ Dr.\ Karl Remeis-Observatory and Erlangen Centre for
  Astroparticle Physics, Sternwartstr.~7, 96049 Bamberg, Germany\\
  $^{2}$ Astronomical Institute, Academy of Sciences,
  Bo\v{c}n\'{\i}~II 1401, CZ-141\,31~Prague, Czech
  Republic\\
  $^{3}$ XMM-Newton Science Operations Centre, ESA, Villafranca del
  Castillo,
  Apartado 78, 28691 Villanueva de la Ca{\~nada}, Spain \\
  $^{4}$ Kavli Institute for Astrophysics and Space Research,
  Massachusetts Institute of Technology, 77 Massachusetts Avenue,
  Cambridge, MA 02139, USA }

\begin{document}


\pagerange{\pageref{firstpage}--\pageref{lastpage}} \pubyear{2011}

\maketitle

\label{firstpage}

\begin{abstract}
  We present the results of a 500\,ksec long \textsl{XMM-Newton}
  observation and a 120\,ksec long quasi-simultaneous \textsl{Chandra}
  observation of the Narrow Line Seyfert 1 galaxy 1H0707$-$495
  performed in 2010 September. Consistent with earlier results by
  \citet{Fabian2009a} and \citet{Zoghbi2010a}, the spectrum is found
  to be dominated by relativistically broadened reflection features
  from an ionised accretion disc around a maximally rotating black
  hole. Even though the spectra changed between this observation and
  earlier \textsl{XMM-Newton} observations, the physical parameters of
  the black hole and accretion disc (i.e., spin and inclination) are
  consistent between both observations. We show that this reflection
  spectrum is slightly modified by absorption in a mildly
  relativistic, highly ionised outflow which changed velocity from
  around $0.11\,c$ to $0.18\,c$ between 2008 January and 2010
  September. Alternative models, in which the spectral shape is
  dominated by absorption, lead to spectral fits of similar quality,
  however, the parameters inferred for the putative absorber are
  unphysical.
\end{abstract}


\section{Introduction}

The Narrow Line Seyfert 1 galaxy 1H0707$-$495 ($z=0.04057$,
\citealt{Jones2009aMNRAS}) is most famous for its huge and sharp drop
in flux at $\sim7\,$keV. Discovered by \citet{Boller2002a}, this
feature was initially interpreted as an absorption edge despite the
lack of a fluorescent emission line. Subsequent models for the feature
included partial covering by a thick absorber \citep{Tanaka2004a},
which also explains the temporal changes seen in its location
\citep{Gallo2004a}. In a long \textsl{XMM-Newton} observation,
however, \citet{Fabian2009a} and \citet{Zoghbi2010a} showed that
relativistically broadened fluorescent iron K$\alpha$ emission is a
more likely explanation, especially given that these authors could
also show that the source's soft-excess can be modelled by a
relativistic iron L$\alpha$ line, the first one ever observed. That
the position of the two lines and the ratio of the normalisations
agree with expectations adds further evidence to the reflection
scenario \citep{Fabian2009a}. In order to produce such a strong
feature and an observable Fe L$\alpha$ line, the iron abundance in
1H0707$-$495 has to be several times the Solar composition. The new
observations also ruled out the partial covering explanation for the
feature, as the strong absorption lines and edges predicted by these
models are not compatible with high resolution X-ray spectra
\citep{Zoghbi2010a}. Recently, \citet{Fabian2012a} observed
1H0707$-$495 at low flux, confirming the reflection dominated
interpretation. The same physical interpretation also successfully
describes the spectra of similar AGN, e.g., IRAS13224$-$3809
\citep{Boller2003a,Ponti2010a}, too.

In this \emph{paper} we analyse data of 1H0707$-$495 taken by the
\textsl{XMM-Newton} satellite from 2010 September 12 until 2010
September 19. The results are compared to a previous observation
\citep{Fabian2009a,Zoghbi2010a} with similar exposure and flux.
Section~\ref{sec:data-analysis} is dedicated to present the data
reduction we applied and general properties of the observation are
discussed. The outcome of our spectral modelling is presented in
Sect.~\ref{sec:x-ray-spectrum} and finally the results are
discussed in Sect.~\ref{sec:results}.

\section{Data Analysis}
\label{sec:data-analysis}

We analyse data of 1H0707$-$495 from a $\sim$500\,ksec long observing
run of \textsl{XMM-Newton} satellite \citep{Jansen2001a} from 2010
September 12 until 2010 September 19 (\textsl{XMM-Newton} revolutions
1971--1974, corresponding to the Obs. IDs 0653510301, 0653510401,
06535105010, and 653510601). These data are then compared to
$\sim$500\,ksec of measurements from an earlier \textsl{XMM-Newton}
observation of the source performed from 2008 January 29 to 2008
February 06. We concentrate on data from the EPIC-pn camera
\citep{Strueder2001aMNRAS} and the reflection grating spectrometer
\citep[RGS;][]{Herder2001aMNRAS}. Data were reduced using the
\textsl{XMM-Newton} Software Analysis System (SAS v.11.0.0) and the
newest calibration files.

Both sets of observations were analysed in the same way. We merged the
linearised event files of the four consecutive observations to a
single event file. These data were cleaned for high background times
following \citet{schartel:07a} and \citet{piconcelli:05a}, resulting
in final exposure times of 328\,ksec and 410\,ksec for \ro\ and \rn,
respectively. The EPIC-pn source spectra were extracted from a
circular region with a radius of $36''$ centred on the maximum of
source emission. In order to avoid any issues due to background Cu K
emission lines from the electronic circuits on the back side of the
detector \citep[see][]{Zoghbi2010a}, background data were taken from a
different chip, using a circular region of the same size at a CCD
position located at a similar distance from the readout node as the
source position (see ``XMM-Newton Users Handbook'', Issue 2.9, 2011,
ESA: XMM-Newton SOC).

\begin{figure}
  \includegraphics[width=\columnwidth]{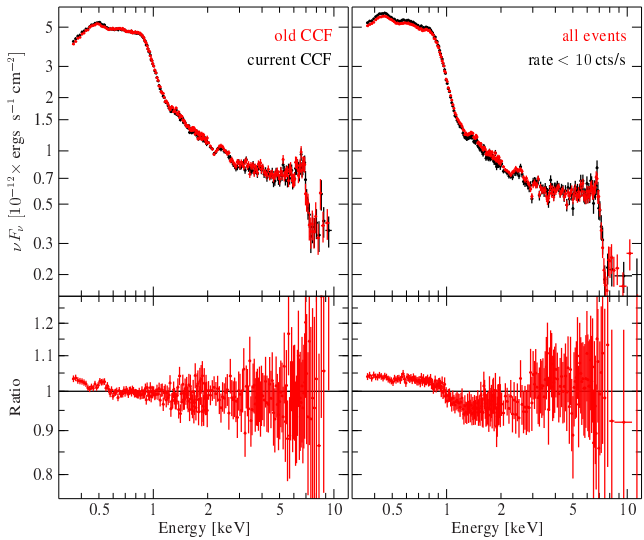}
  \caption{\textbf{Left}: Comparison of the extractions for \ro\
    between the current calibration of the EPIC-pn (black, calibration
    date 2011-09-13) and the older version (red, 2009-07-01). The
    bottom figure shows the ratio between the data sets.
    \textbf{Right}: Spectra of \rn\ obtained for a ``normal''
    extraction (red) and an extraction with the conditions
    \mbox{\texttt{RATE $< 10$}} (black). All spectra are normalised in
    flux to the ``black'' spectra.}
  \label{fig:pile-up}
\end{figure}

In order to exclude any contamination of the spectra due to pile up, a
light curve with a resolution of 40\,s was generated and only times
with count rates below $10\,\mathrm{counts}\,\mathrm{s}^{-1}$ were
included in the final spectrum (SAS extraction expression
\texttt{\mbox{(RATE$ > 0.5$)} \&\& \mbox{(RATE$\le 10.0$)} \&\&
  \mbox{(FRACEXP$> 0.7$)}}, the last expression guarantees that each
bin has an exposure of at least 70\%)\footnote{This selection is not
  fulfilled for 2.4\% (\ro) and 5.3\% (\rn) of these time bins.}.
Excluding data with count rates above
$10\,\mathrm{counts}\,\mathrm{s}^{-1}$ changes the average flux by
about 5\% (Fig.~\ref{fig:pile-up}, right). These changes are highest
in the lower energy band, which dominates the overall source spectrum
and is therefore the spectral part which is crucial for determining
the spectral parameters.
The light curves were extracted in the same manner, except that we
dropped the condition \mbox{(RATE$\le 10.0$)}.

We note that since the publication of the earlier \textsl{XMM-Newton}
observation of 1H0707$-$495 \citep{Fabian2009a,Zoghbi2010a} the
calibration of the EPIC-pn camera has significantly changed in the
soft X-rays (3\% below 0.6\,keV, see Fig.~\ref{fig:pile-up},
left). The newer response is now in line with simultaneous
measurements from the RGS (see XMM-CCF-REL-266, available at
\url{http://xmm2.esac.esa.int/external/xmm_sw_cal/calib/rel_notes/index.shtml}).
In order to take into account any remaining systematic uncertainties
in the complex spectral models as well as any remaining calibration
uncertainty, a systematic error of 3\% is added to data taken below
1.2\,keV.

The RGS spectra were processed with the SAS task \texttt{rgsproc} with
calibration files created mid April 2011. The observations were
screened for high background times in the standard way and run with
the RGS rectification on (see XMM-CCF-REL-269, available at
\url{http://xmm2.esac.esa.int/external/xmm_sw_cal/calib/rel_notes/index.shtml}),
which allows a direct comparison with EPIC-pn results.

Simultaneously to the \textsl{XMM-Newton} observations, between 2010
September 12 and 2010 September 19, we also observed 1H0707$-$495 with
the \textsl{Chandra} satellite \citep{canizares:05a} for a total of
118.2\,ksec (Obs. ID 12115, 12116, 12117, and 12118). The Medium and
High Energy Transmission Gratings spectra (METG and HETG) were
extracted and the grating ARFs and RMFs were created using the
standard \texttt{ciao} threads (v4.1) and \texttt{caldb} v4.1.2. The
spectra for each observation and for the +1 and -1 orders were then
combined using standard \texttt{ciao} scripts.

Spectral fitting was performed with the \textsl{Interactive Spectral
  Interpretation System} \citep[\textsl{ISIS}; ][]{Houck2000a}.  Data
were rebinned to oversample the intrinsic energy resolution of the
EPIC-pn camera slightly, requiring a rebinning to 2, 3, 10, 15, and 25
bins for energies above 0.8, 2.0, 4.0, and 7.0\,keV, respectively,
where one bin has a width of 5\,eV.  Where necessary, data were
rebinned further in order to reach a signal to noise ratio of $S/N >
2.5$.  All uncertainties are given at 90\% confidence if not stated
differently.

\section{The X-ray spectrum of 1H0707$-$495}
\label{sec:x-ray-spectrum}
\subsection{Introduction}

\begin{figure}
  \includegraphics[width=\columnwidth]{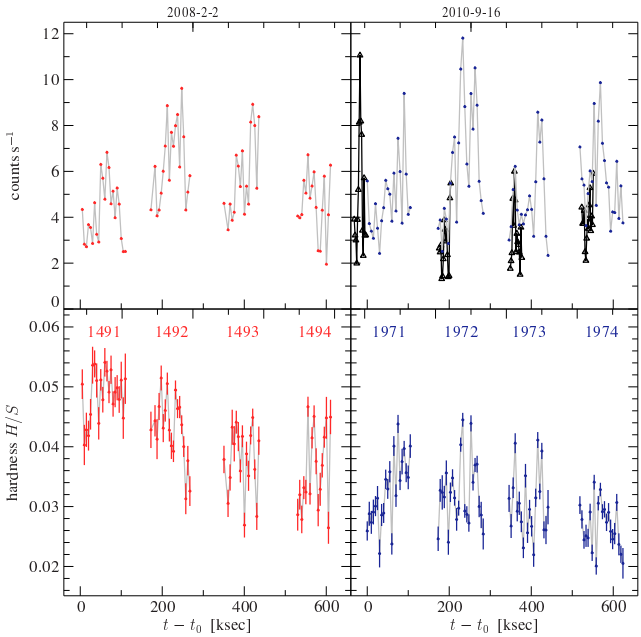}
  \caption{\textbf{Left:} Light curve of the measurements analysed by
    \citet{Zoghbi2010a} (Revs.\ 1491--1494, red) and \textbf{Right:}
    Light curve from the 2010 September observations (Revs.\
    1971--1974, blue). Both light curves are binned to a time
    resolution of 5\,ksec for the energy band 0.3--10.0\,keV. In
    addition, the simultaneous \textsl{Chandra} light curve for
    0.35--8.0\,keV is shown below the \textsl{XMM-Newton} light curve
    (black, triangles). Due to lower count rate, the rate was
    multiplied by a factor of 50 for comparison. \textbf{Bottom
      panels:} X-ray hardness for both observations. $S$ and $H$
    denote source counts in the bands 0.35--1.5\,keV and
    2.0--8.0\,keV, respectively.}
  \label{fig:lightcurve}
\end{figure}

\begin{figure}
  \includegraphics[width=\columnwidth]{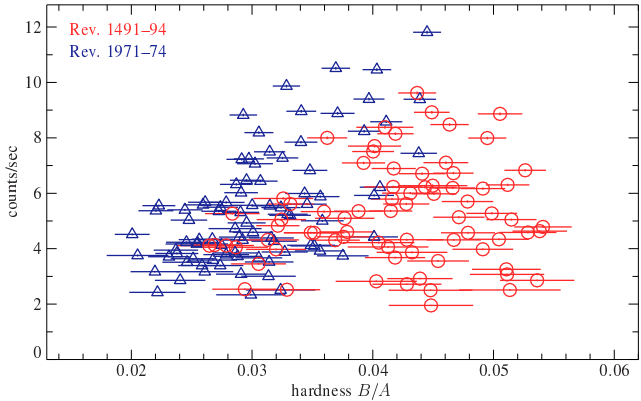}
  \caption{Hardness-intensity diagram of the observations, based on a
    time resolution of 5\,ksec. The new observation (\rn, blue
    triangles) is slightly but significantly softer than \ro\ (red
    circles). There is a positive correlation between the hardness and
    the intensity, which is stronger for \rn\ (correlation coefficient
    $\rho = 0.60$) than for \ro\ ($\rho = 0.23$).}
  \label{fig:hardness}
\end{figure}

1H0707$-$495 is known from previous observations to be a highly
variable Narrow Line Seyfert 1 Galaxy \citep{leighly:99a,turner:99a}.
Figure~\ref{fig:lightcurve} shows the light curve and hardness ratio
during the observation in \ro\ and \rn. In comparison the older
observation (\ro) was a little less variable (rms variability of 6\%
compared to 9\% for 100\,s bins) and slightly weaker
(4.6\,cts\,s$^{-1}$ in average compared to 5.1\,cts\,s$^{-1}$). In
combination with the hardness-intensity diagram
(Fig.~\ref{fig:hardness}), where no distinct difference between the
two observations can be seen, it is obvious that the source was in a
similar state. Nevertheless, this diagram reveals that data from the
newer observation are slightly softer (mean hardness ratio of 0.030
compared to 0.038). In Sect.~\ref{sec:basic-model} we will show that
the softening originates from a softer power law index and a stronger
soft-excess, but does not influence other fitting parameters
significantly.

\begin{figure}
  \includegraphics[width=\columnwidth]{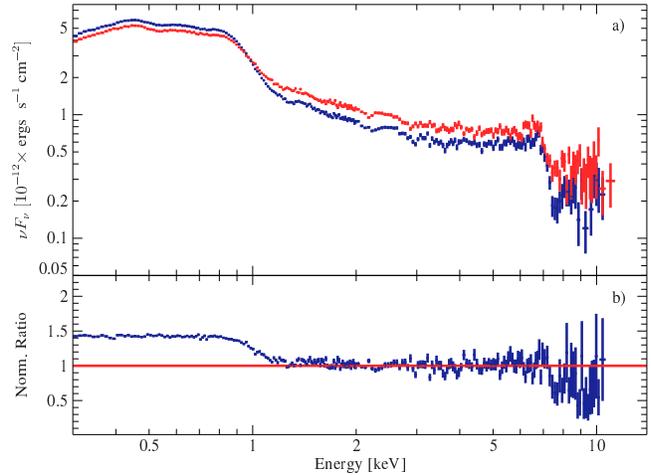}
  \caption{(a) Unfolded spectra of \ro\ (red) and \rn\ (dark
    blue). The bottom panel (b) shows the ratio of the unfolded
    spectra (\rn\ divided by \ro). It demonstrates that the new
    observation becomes softer, but most spectral features remain the
    same. For illustrative purposes the ratio spectrum was multiplied
    by a constant factor such that the 2--5\,keV range coincides with
    unity.}
  \label{fig:compare_obs}
\end{figure}

Figure~\ref{fig:compare_obs} compares the unfolded spectra of \ro\ and
\rn. Apart from the overall similar spectral shape, a large number of
the smaller spectral features are similar, including the ``wiggles''
in the 2--5\,keV band that have been attributed to the complex
emission and absorption spectrum of the source
\citep{Blustin2009a}. Moreover, the characteristic drop at
$\sim$7\,keV is observed at the same energy. At energies above this
spectral drop, more flux is missing than can be explained by a simple
softening of the source. As will be shown in
Sect.~\ref{sec:highly-ioniz-outfl}, this difference might be due to
variability of an ultra-fast and highly ionised wind, which acts as an
absorber.

\subsection{Broad Band Spectrum and Absorption}
\label{sec:broad-band-and-absorption}

\begin{figure}
  \includegraphics[width=\columnwidth]{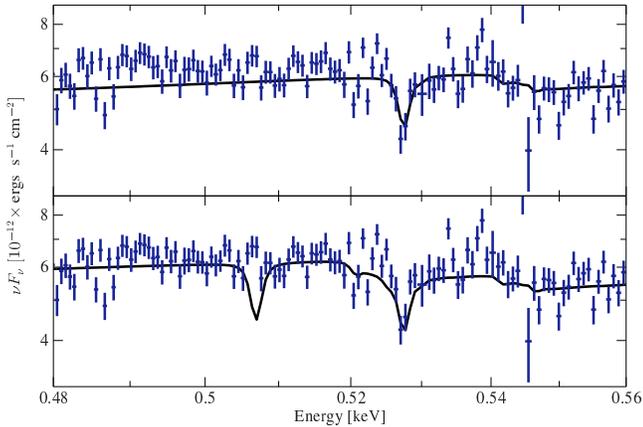}
  \caption{ The EPIC-pn best-fit applied to the RGS spectra of \rn\
    and focusing on the \ion{O}{i} (0.527\,keV) absorption line. The
    upper panel shows the best-fit model for Galactic absorption
    only. The absorption line is clearly seen in the data. Residuals
    from a better fit including source intrinsic absorption
    ($z=0.04057$) are shown in the lower panel. The lack of a
    redshifted \ion{O}{i} line at 0.506\,keV in the measured spectrum
    shows that the additional neutral absorption is not source
    intrinsic.}
  \label{fig:absorption_rgs}
\end{figure}

We model the average X-ray continuum using the standard spectrum for
Narrow Line Seyfert~1 spectra, namely, a steep, absorbed power law
spectrum plus (relativistically smeared) X-ray reflection
\citep[e.g.,][]{lamura:11a,vaughan:99a}. Initially, foreground X-ray
absorption was modelled assuming a equivalent column of $N_\mathrm{H}
= 4\times10^{20}\,\mathrm{cm}^{-2}$ as obtained from the full
resolution data of the Leiden-Argentine-Bonn 21\,cm survey \citep[LAB
survey,][]{Kalberla2005a} and assuming the abundances of
\citet{Wilms2000a}. After adding a weak soft excess modelled by a disk
blackbody \citep{Mitsuda1984a,Makishima1986a} with a temperature of
$kT_\mathrm{in} \sim 100\,$eV, we obtain a basic model which describes
the EPIC-pn data well, although below 0.5\,keV, some residuals
remain. One explanation for these residuals is additional source
intrinsic absorption at a level of $\sim
3\times10^{20}\,\mathrm{cm}^2$. An increased column in this range is
necessary to obtain any good fit at all when extending the lower end
of the spectrum to 0.3\,keV. If true, however, at this column a
resonant \ion{O}{i} K$\alpha$ line at 0.527\,keV in the rest frame of
the source should be visible, i.e., at 0.506\,keV in the measured
spectrum, which is not the case (Fig.~\ref{fig:absorption_rgs}). A
systematic grid search for any other narrow line associated with a
potential absorber failed as well. Any increased neutral
$N_\mathrm{H}$ must therefore be of Galactic origin. We note that like
most 21\,cm $N_\mathrm{H}$ values quoted in X-ray astronomy, the
$N_\mathrm{H}$ value quoted above is from an all-sky 21\,cm survey
with a rather coarse angular resolution of $0.6^\circ$ such that small
scale variations of $N_\mathrm{H}$ are washed out. Within $3^\circ$ of
1H0707$-$495, the LAB-survey contains points with $N_\mathrm{H}$ as
large as $8\times10^{20}\,\mathrm{cm}^{-2}$. In addition, 21\,cm data
only probe the gas phase of the interstellar medium, while a
significant amount of X-ray absorbing material could also be in
molecules. It is not uncommon for higher Galactic latitudes that the
gas and molecular columns are comparable
($N(\mathrm{H}_2)/N(\mathrm{H})$ varies between 0.2 and 5 in the
sample of \citealt{magnani:85a}). Finally, it cannot be ruled out that
some of the excess absorption is source intrinsic and either in mildly
ionised material or in material that is fast enough that narrow
features are smeared out as suggested by \citet{Zoghbi2010a}. We
therefore leave $N_\mathrm{H}$ as a free parameter in our spectral
modelling and do not speculate on the relative fraction of
source-intrinsic and Galactic absorption along the line of sight to
1H0707$-$495 from these fits.

\subsection{Reflection Model}
\label{sec:basic-model}

For the detailed modelling of the spectrum we started with the single
reflection model as proposed as best-fit model by \citet{Fabian2009a}
and \citet{Zoghbi2010a}. This model consists of ionised reflection
from an accretion disc \citep[\texttt{reflionx},
][]{Ross2005a,ross2007a}, which is relativistically smeared using the
\texttt{relconv} model \citep{Dauser2010a}. In addition, a weak disk
blackbody with $kT_\mathrm{in} \sim 100\,$eV was added to describe
the soft-excess.

In this model, relativistic smearing is parametrised using the
emissivity law $\epsilon(r)$ of the accretion disc, which describes
the radius-dependent reflected power of the accretion disc per unit
area \citep[][and references therein]{Dauser2010a}. Initially,
$\epsilon(r)$ was described by a broken power law. Extensive spectral
analysis showed, however, that the parameters for describing such an
emissivity, two spectral indices and the break radius, were only
poorly constrained when allowed to vary freely. As suggested for
1H0707$-$495 by \citet{Wilkins2011a} and \citet{Fabian2012a} and for
Narrow Line Seyfert~1 Galaxies in general by \citet{ghisellini:04a},
we therefore constrain the emissivity to be due to the lamp post
geometry \citep{Martocchia1996a}. In this accretion geometry, the
X-ray continuum is produced by a compact and central source above the
black hole which focuses (part of) the radiation onto the accretion
disc, which extends from the innermost stable circular orbit out to
400\,$r_\mathrm{g}$, where $r_\mathrm{g} = GM/c^2$ is the
gravitational radius. In this model, the only free parameter
determining the emissivity profile is the height of the point-like
source of X-rays above the black hole. As the emissivity in the lamp
post geometry can also be approximated by a broken power law to
$0^\mathrm{th}$ order \citep[see, e.g., Fig.~11 of ][]{Fabian2012a},
we were easily able to reproduce any good fit in the lamp post
geometry with an equally good one for a broken power law emissivity.

\begin{figure*}
  \includegraphics[width=\textwidth]{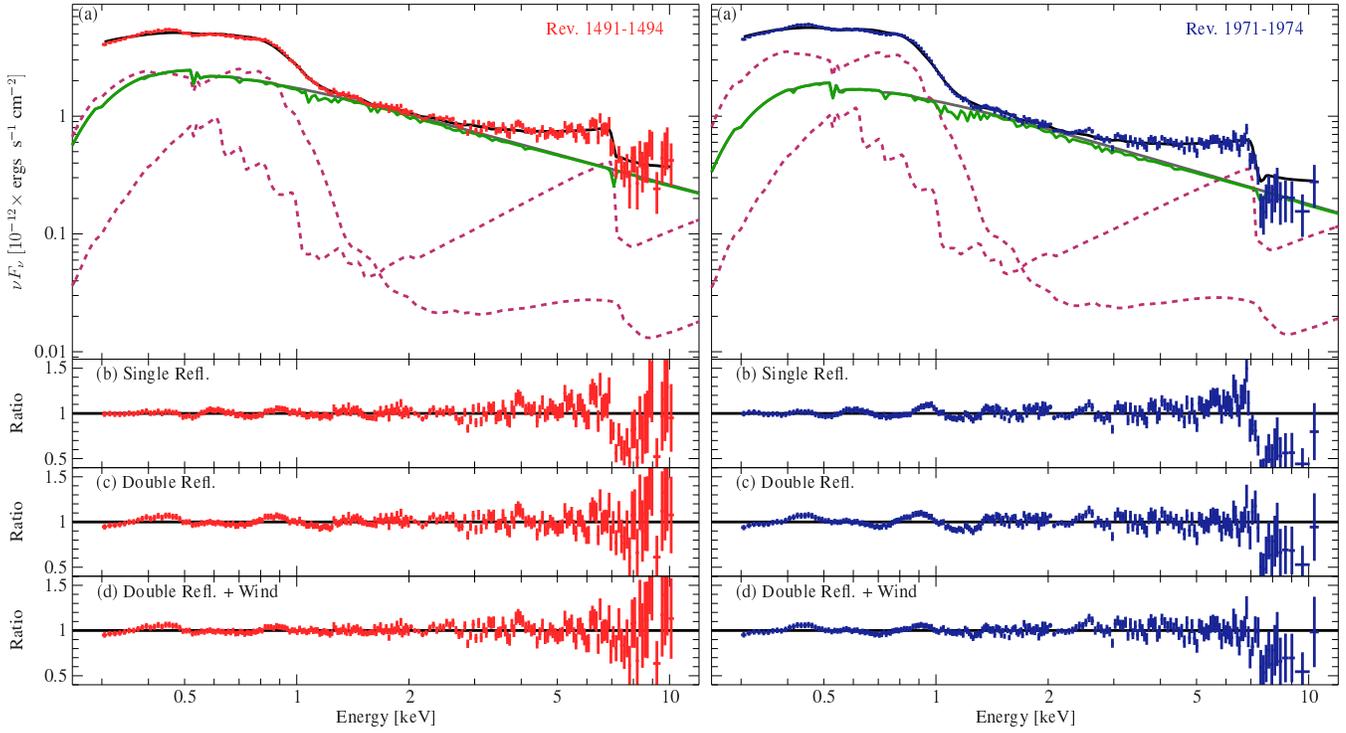}
  \caption{Best-fit model of \ro\ (left) and \rn\ (right). (a) The
    upper panel shows the best-fit model components: Two relativistic
    reflection components (purple, dashed) and a power law (dark
    gray), absorbed by a highly ionised and blue-shifted outflow
    (green). The lower panels show the residuals of (b) a single
    reflection component, (c) two reflection components, and (d) the
    best-fit model with additional absorption by an ionised wind.}
  \label{fig:spectra}
\end{figure*}

This model is capable of roughly describing the data
($\chi^2/\mathrm{dof} = 359/220 \sim 1.63$ and $\chi^2/\mathrm{dof}=
503/214 \sim 2.35$ for \ro\ and \rn, respectively), with relativistic
parameters for both observations being similar to \citet{Zoghbi2010a}.
Figure~\ref{fig:spectra}b shows the corresponding fit and residuals,
Table~\ref{tab:table} lists the best-fit parameters. We note that the
best-fit values of height and inclination for \ro differ significantly
from those values obtained in the other models. However, there exists
some degeneracy between the best-fit values for these parameters, as
the model only roughly describes the data. Therefore the fit is found
to be almost as good, when the height and inclination are fixed to the
best-fit values obtained for the other fits.

Especially for \rn, the sharp drop in the spectrum at $\sim$7\,keV is
not well modelled by a single reflection. Using a second reflection
component, which is highly ionised, significantly improves the fit
($\Delta \chi^2 (\mbox{\rn})= 64$, Fig.~\ref{fig:spectra}c). This
second reflection component acts as an additional layer on top of the
moderately ionised component, i.e., the reflectors are not radially
separated. In this composite model, emission from the highly ionised
reflection component at low energies describes the soft excess, i.e.,
an additional blackbody component is not required, while the low
ionised reflection with strong K$\alpha$ emission line accounts
properly for the strong drop in flux around 7\,keV.

Even in this two component reflection fit, however, some residuals
remain in the area around $\sim$1\,keV, especially for \rn. These
residuals can be described by a highly ionised ($\log(\xi)\sim3.5$)
and smeared outflow, modelled here using the \texttt{swind} model
\citep[][see Fig.~\ref{fig:spectra}d and green line in
Fig.~\ref{fig:spectra}a]{Gierlinski2004a}. Particularly for \rn, such
an absorption improves the fit a lot ($\Delta\chi^2 = 165 $).
Moreover, the blue-shifted absorption of \ion{Fe}{xxv} and
\ion{Fe}{xxvi} at 6.7\,keV and 7.0\,keV is automatically predicted at
the position of the prominent absorption feature at $\sim$7.5\,keV
seen directly below the sharp drop. The \texttt{swind} model describes
the remaining residuals in a satisfactory manner, with outflow
velocities of $0.18\pm0.01\,c$ and $0.11\pm^{+0.01}_{-0.02}\,c$ for
\rn\ and \ro, respectively, calculated from the determined redshift
parameter $z_\mathtt{swind}$ (see Tab.~\ref{tab:table}). Winds at
these speeds are commonly seen in many sources \citep[see, e.g.,
][]{Tombesi2010a}. We will discuss this result in greater detail below
(Sect.~\ref{sec:highly-ioniz-outfl}). In addition, the smearing due to
a turbulent velocity is in the order of $0.01\,c$--$0.02\,c$, i.e.,
3000--$6000\,\mathrm{km}\,\mathrm{s}^{-1}$, which is in agreement with
the gratings spectrum (see Sect.~\ref{sec:broad-band-and-absorption}).
Despite this overall success of the model, however, the predicted
strength of these lines is too weak. This weakness is a limitation of
the \texttt{swind} model, which is constrained to material of Solar
abundances and does not allow us to model these lines using the
extreme metallicity indicated by the reflection component.

Despite the large uncertainties, at each point above 7\,keV in \rn\
there is a systematic over-prediction of the flux (see
Fig.~\ref{fig:spectra}d, right). It probably originates from improperly
modelled accretion disk reflection; in particular, contributions from
the lower ionised component contribute to a large amount of the total
flux observed in this energy range (see Fig.~\ref{fig:spectra}a). As
will be discussed in detail in Sect.~\ref{sec:results}, the modelling
of the reflection suffers from several constraints. For example, the
simplified modelling of the accretion disk spectrum, or the fact that
Fe is highly over-abundant while all other elements are at Solar
abundance, could easily produce an over-prediction of the model flux
above 7\,keV. In addition to that, the mildly ionised wind we find can
only be modelled assuming Solar composition, too, although we conclude
from the reflection component that Fe is highly over-abundant. We
therefore expect that our model under-predicts the absorption by
H-like and He-like transitions of iron in the energy range around
7\,keV.

\begin{table*}
  \caption{Best-fit models, as shown in Fig.~\ref{fig:spectra}:
    Model~1 is a power law, a single broadened reflection component and a
    blackbody. Model~2 consists of two reflection components of
    different ionisation parameter, $\xi$, and Model~3 is the best-fit
    model, based on Model~2 combined with absorption of radiation in a
    highly ionised wind. The ionisation parameter is defined as $\xi =
    4\pi F/n_\mathrm{e}$,
    with flux, $F$, and electron number density, $n_\mathrm{e}$. The
    parameter $v_\mathrm{swind}$ is the velocity dispersion of the
    wind and $z_\mathrm{swind}$ its redshift compared to the rest
    frame. Note that in most models, the iron abundance relative to
    Solar in the \texttt{reflionx} component, $Z_\mathrm{Fe}$, pegs at
    the upper limit of $Z_\mathrm{Fe} = 20$, while all other elemental
    abundances are fixed to Solar.
  }
  \label{tab:table}   
  {
    \footnotesize
    \begin{tabular}{l| c  c  c  c  c  c }
\hline\hline  & \multicolumn{2}{c}{ \textbf{Model 1} (Single Refl.)} &\multicolumn{2}{c}{ \textbf{Model 2} (Double Refl.)} &\multicolumn{2}{c}{ \textbf{Model 3} (Double Refl. + Wind)} \\
 & \ro  & \rn  & \ro  & \rn  & \ro  & \rn \\ \hline
$A_\Gamma \times 10^{-3} $ & $0.958^{+0.017}_{-0.029}$ & $0.727^{+0.013}_{-0.014}$ & $1.093^{+0.023}_{-0.031}$ & $0.836^{+0.018}_{-0.020}$ & $1.182^{+0.030}_{-0.027}$ & $0.913^{+0.024}_{-0.025}$ \\
$\Gamma$ & $2.575^{+0.032}_{-0.022}$ & $2.609^{+0.025}_{-0.020}$ & $2.86^{+0.04}_{-0.05}$ & $2.91\pm0.04$ & $2.86^{+0.05}_{-0.04}$ & $2.853^{+0.027}_{-0.025}$ \\
\hline  $A_{\mathrm{refl}_1} \times 10^{-5}$ & $0.096^{+0.032}_{-0.025}$ & $0.088^{+0.026}_{-0.016}$ & $2.0^{+0.6}_{-0.5}$ & $2.7^{+0.7}_{-0.4}$ & $3.6^{+0.4}_{-1.4}$ & $4.8^{+0.6}_{-2.0}$ \\
$A_{\mathrm{refl}_2} \times 10^{-5}$ & - & - & $0.0058^{+0.0018}_{-0.0010}$ & $0.0111^{+0.0013}_{-0.0010}$ & $0.0056^{+0.0010}_{-0.0007}$ & $0.0048^{+0.0013}_{-0.0008}$ \\
$ \log ( \xi_{\mathrm{Refl}_1} )$ & $1.965^{+0.059}_{-0.030}$ & $2.074^{+0.027}_{-0.022}$ & $1.301^{+0.014}_{-0.148}$ & $1.301^{+0.010}_{-0.109}$ & $1.00^{+0.14}_{-0.00}$ & $1.004^{+0.251}_{-0.004}$ \\
$ \log ( \xi_{\mathrm{Refl}_2} )$ & - & - & $2.995^{+0.012}_{-0.115}$ & $3.000^{+0.004}_{-0.051}$ & $3.000^{+0.027}_{-0.081}$ & $3.301^{+0.006}_{-0.118}$ \\
$Z_\mathrm{Fe}$ & $19.73^{+0.28}_{-11.54}$ & $20.0^{+0.0}_{-1.0}$ & $10.3^{+3.4}_{-2.5}$ & $12.0^{+2.8}_{-2.5}$ & $15\pm5$ & $20^{+0}_{-6}$ \\
\hline  $h^\mathrm{lp}\;[r_\mathrm{g}]$ & $29^{+8}_{-5}$ & $3.0^{+0.6}_{-0.0}$ & $3.13^{+0.71}_{-0.13}$ & $3.00^{+0.28}_{-0.00}$ & $3.0006^{+0.3372}_{-0.0006}$ & $3.00^{+0.13}_{-0.00}$ \\
$a^\mathrm{lp}$ & $1.0^{+0.0}_{-1.2}$ & $0.971^{+0.016}_{-0.017}$ & $0.998^{+0.000}_{-0.009}$ & $0.9980^{+0.0000}_{-0.0023}$ & $0.998^{+0.000}_{-0.008}$ & $0.998^{+0.000}_{-0.011}$ \\
$\theta^\mathrm{lp}$ [deg] & $77.7^{+2.3}_{-4.8}$ & $58.4^{+0.9}_{-1.3}$ & $48.8^{+1.4}_{-1.9}$ & $48.8\pm1.0$ & $52.0^{+1.7}_{-1.8}$ & $48.8^{+1.3}_{-1.2}$ \\
\hline  $N_\mathrm{H}\;[10^{22}\,\mathrm{cm}^{-2}]$ & $0.070^{+0.019}_{-0.009}$ & $0.082\pm0.010$ & $0.052^{+0.004}_{-0.005}$ & $0.0589^{+0.0024}_{-0.0025}$ & $0.0545^{+0.0027}_{-0.0030}$ & $0.0628^{+0.0025}_{-0.0028}$ \\
$A_\mathrm{bb} \times 10^4$ & $0.59^{+1.07}_{-0.19}$ & $1.7^{+1.0}_{-0.7}$ & - & - & - & - \\
k$T_\mathrm{bb}$ [eV] & $112^{+8}_{-17}$ & $96^{+7}_{-6}$ & - & - & - & - \\
$N_\mathrm{H}^\mathrm{swind} \;[10^{22}\,\mathrm{cm}^{-2}]$ & - & - & - & - & $3.6^{+7.0}_{-0.6}$ & $4.0^{+1.3}_{-1.0}$ \\
$\log(\xi_\mathrm{swind})$ & - & - & - & - & $3.61^{+0.17}_{-0.07}$ & $3.299^{+0.024}_{-0.071}$ \\
$v_\mathrm{swind}\;[c]$ & - & - & - & - & $\le0.025$ & $\le0.014$ \\
$z_\mathrm{swind}$ & - & - & - & - & $-0.063^{+0.015}_{-0.007}$ & $-0.1262^{+0.0015}_{-0.0036}$ \\
\hline
$\chi^2/\mathrm{dof}$  & $359/220 = 1.63 $ & $503/214 = 2.35 $ & $290/220 = 1.32 $ & $439/214 = 2.05 $ & $237/216 = 1.10 $ & $274/210 = 1.31 $ \\
\hline\hline \end{tabular}

  }
\end{table*}

\subsection{Alternative Models: Smeared Absorption}
\label{sec:altern-models:-smear}

\begin{figure}
  \includegraphics[width=\columnwidth]{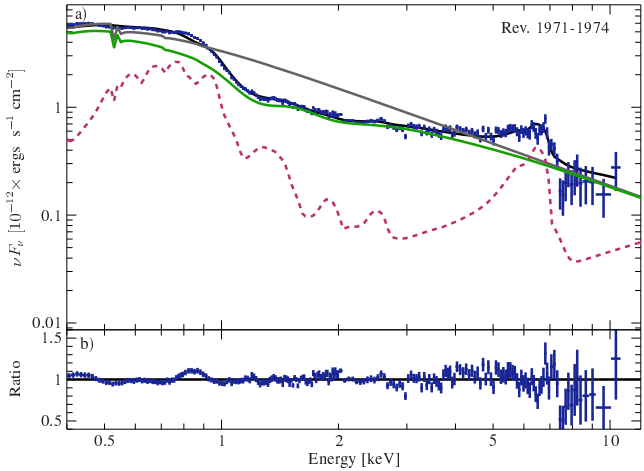}
  \caption{An alternative model (here for \rn) which describes the
    soft-excess as a blend of several mildly ionised
    ($\log(\xi)\sim2.2$) and smeared Oxygen absorption lines. (a)
    Single components: The effect of the absorption on the original
    power law (dark gray) is illustrated by the green line and the
    reflection component is shown by the purple, dashed line. (b)
    Residuals of the best-fit ($\chi^2/\mathrm{dof} = 375/226 \sim
    1.66$). }
  \label{fig:alter_spectra}
\end{figure}

As an alternative to the ``pure'' reflection model, we also tried to
model the soft-excess by a strongly ionised and ultra-fast outflow in
a manner proposed by \citet{Gierlinski2004a}. As discussed in
Sect.~\ref{sec:broad-band-and-absorption}, there are no narrow lines
present in the RGS spectrum of 1H0707$-$495. We therefore use again
the \texttt{swind} model to describe smeared absorption.
Figure~\ref{fig:alter_spectra} shows that this model together with a
single reflection component is capable of describing the
data. Compared to the residuals in Fig.~\ref{fig:spectra}, however,
the emission-like hump around 0.9\,keV is much more pronounced. In
order to work, the model requires an outflow velocity of $\sim0.37\,c$
of the absorbing material, which is larger than most relativistic
outflows observed in Active Galaxies \citep[see,
e.g.,][]{Blustin2005a,Tombesi2010a}. More doubtful is that in order to
explain the data a line broadening with a rms velocity distribution of
as high as $50000\,\mathrm{km}\,\mathrm{s}^{-1}$ has to be
assumed. From a statistical point of view, however, this model cannot
be completely rejected ($\chi^2/\mathrm{dof} = 375/226 \sim 1.66 $ for
\rn).

Finally, as argued by \citet{Zoghbi2010a}, if the spectral drop was
due to a partially covering absorber a strong neutral K-line would be
expected. For the upper limit search we use the summed
\textsl{Chandra} HETG spectrum without any further binning. With these
observations we are able to determine the upper limit of the flux of a
narrow ($\sigma=1$\,eV) Gaussian line at 6.4\,keV to $F_{6.4} <
2.0\times10^{-6}\,\mathrm{photons}\,\mathrm{s}^{-1}\,\mathrm{cm}^{-2}$
at 90\% confidence. This upper limit is consistent with our EPIC-pn
spectral modelling, but does not allow to constrain the modelling
further. The absence of this line again argues against the absorption
interpretation.

\subsection{A highly ionised outflow in 1H0707$-$495}
\label{sec:highly-ioniz-outfl}

\begin{figure}
  \centering
  \includegraphics[width=\columnwidth]{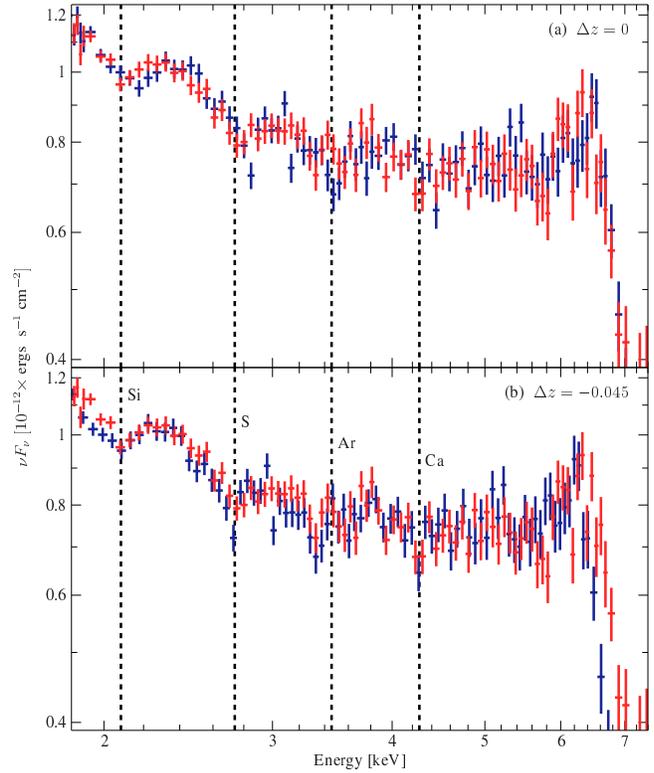}
  \caption{(a) A zoom into the hard X-ray bandpass of the spectra: For
    illustrative purposes \rn\ (dark blue) is renormalised to the same
    $\nu F_\nu$ in the 2--5\,keV band as \ro\ (red). Energies are
    given in the rest frame of the source. (b) Here we apply a manual
    energy shift to \rn\ (blue) corresponding to a redshift of $\Delta
    z = z_{\mathrm{Rev.}\,19} - z_{\mathrm{Rev.}\,14} = -0.045$, which
    leads to the best agreement for most absorption-like features that
    are seen consistently in both observations. The most prominent of
    these features are indicated by dashed lines. Assuming an outflow
    velocity in \ro\ of $\sim$0.04\,$c$ they can be attributed to
    1s-2p transitions of H-like Si, S, Ar, and Ca.}
  \label{fig:compare_obs_zoom}
\end{figure}

A highly ionised and fast outflow was already used by
\citet{done2007a} to describe the spectrum of 1H0707$-$495, where
strong absorption in this wind models the complete spectral shape,
including the sharp drop at $\sim$7\,keV as a P Cygni line. In
contrast, in our best-fit model the main spectral features (the
soft-excess and the sharp drop) are described by reflection, while the
wind, being more ionised, has a smaller influence on the observed
spectrum. We saw in Sect.~\ref{sec:basic-model} that a major
difference between \ro\ and \rn\ is a difference of $\sim$$0.07\,c$ in
wind velocity. Comparing the two spectra for the higher energetic
range (Fig.~\ref{fig:compare_obs_zoom}a) reveals that of all the
structure seen in this energy band only the steep drop around
$\sim$7\,keV agrees in both observations. This constancy is expected
if the 7\,keV feature is due to a relativistic line emitted from an
accretion disc. If one wanted to explain the full spectrum by
absorption, however, it would be very unlikely that all spectral
features except for the drop change their location and that only the
strongest feature stayed constant.

Taking a closer look at the spectra away from the 7\,keV feature
reveals that indeed the same features appear in both observations, but
they are slightly shifted in energy. Correcting for this shift ``by
eye'' we find that most features agree nicely if the shift is
approximately \mbox{$\Delta z = z_{\mathrm{Rev.}\,19} -
  z_{\mathrm{Rev.}\,14} \sim -0.045$}
(Fig.~\ref{fig:compare_obs_zoom}b). This relative shift is consistent
with the difference in redshift found between the two swind components
obtained by spectral modeling, $\Delta z_\texttt{swind} =
-0.06^{0.01}_{-0.02}$. We therefore attribute the fine spectral structure
between 2--5\,keV to a fast and highly ionised outflow.

Assuming redshifts of $z_{\mathrm{Rev. 14}} = -0.04$ and
$z_{\mathrm{Rev. 19}} = z_{\mathrm{Rev. 14}} + \Delta z = -0.085$, the
structures in the spectrum can tentatively be identified with single
1s-2p transitions of H-like Si, S, and Ca
(Fig.~\ref{fig:compare_obs_zoom}b). Again, the individual redshifts
found by eye agree with the ones found in our spectral fits for the
two swind components, $-0.06^{+0.02}_{-0.01}$ and $-0.13\pm0.01$. In
such a highly ionised plasma we would also expect a strong resonant
transition from Argon. Inspection of the Ar band reveals a feature
that is more complex than the absorption features in the Si, S, and Ca
band, making its interpretation difficult at the resolution of an
X-ray CCD. A possible explanation could be a P Cyg profile of H-like
Ar, however, in this case Ar would have a different redshift than the
other features, which seems unlikely. We note, however, that the
overall shape of this feature remained stable between both
observations. Further modeling of the spectral features using proper
wind models is therefore required, which is outside of the scope of
this paper.

\section{Summary and Discussion}
\label{sec:results}

In this paper we have presented the results from a spectral analysis
of 500\,ksec of new \textsl{XMM-Newton} and \textsl{Chandra} data from
the AGN 1H0707$-$495 and have compared these results to the previous
long observation of the source. In agreement with earlier work
\citep{Fabian2009a,Fabian2012a,Zoghbi2010a}, the simple picture
deduced from these observations is that 1H0707$-$495 an accretion disc
around a maximally rotating black hole is irradiated by a central and
compact source closely above the black hole, producing the underlying
complex X-ray continuum by reflection. In both observations, the
primary X-ray source which irradiates the accretion disc has a height
of $\le 4\,r_\mathrm{g}$ above the black hole. Such a low emitter is
in line with timing measurements \citep{Zoghbi2010a} and implies that
the photons are extremely focused onto the inner parts of the disc.
Fits with an empirical broken power law emissivity confirm this
interpretation ($\epsilon_{r<r_\mathrm{br}} = 10.2^{+0.8}_{-0.9}$,
$\epsilon_{r>r_\mathrm{br}} = 2.3^{+0.5}_{-0.1}$, $r_\mathrm{br} =
2.9^{+0.2}_{-1.0}\,r_\mathrm{g}$ for \rn). Despite significant changes
in the continuum shape between both observations, the black hole in
1H0707$-$495 is consistently well determined to be maximally
rotating. This result also adds further credibility to the
relativistic line interpretation of the spectrum, as the black hole
parameters are not expected to change on such short time scales. Such
a behavior would be significantly more difficult to explain in
alternative models explaining the line via an ionised and blue-shifted
radial wind.

The accretion disc causing the observed reflection features has a
complex ionisation structure, which is here approximated by two
reflectors of strongly different ionization as modeled by
\citet{Ross2005a} and \citet{ross2007a}. Even though the highly
ionised reflector dominates the soft X-ray spectrum, its normalisation
is only 0.1--0.2\% of that of the weakly ionised reflector, i.e., most
of the irradiated X-rays are intercepted by the colder medium. This
result fits nicely into the picture of a rather neutral accretion disc
with a thin skin of highly ionised material caused by the incident
X-rays \cite[see, e.g.,][]{garcia:10a,garcia:11a}. As a caveat,
however, note that our approach implicitly assumes that the ionisation
fraction of the disc is independent of distance from the black hole.
From basic arguments it is clear that the photoionisation due to the
incident X-rays will result in an ionisation structure of the disc
surface which strongly depends on radius, as even in the simplest and
most conservative models photons are strongly focused towards the
inner parts. Additionally, the energy release within the accretion
disc itself will give rise to a temperature and ionisation gradient
\citep[e.g.,][and references therein]{hubeny:01a,davis:05a}.

As shown by our modeling of the weaker spectral features, the
relativistically blurred continuum is then modified by absorption in
an ultra-fast wind ($0.11$--$0.18\,c$, i.e.,
30000--50000\,$\mathrm{km}\,\mathrm{s}^{-1}$). Due to the change of
the minor features at 2--5\,keV and the constancy of the 7\,keV drop
between the two observations (see \ref{sec:highly-ioniz-outfl}), the
ionised wind could be uniquely identified to exist along with the
reflection, which dominates the spectrum. The overall properties of
this wind are in line with the relativistic, highly ionized winds that
have now been detected in more than 40 radio-quiet AGN, including
Narrow Line Seyfert 1 galaxies. These winds have been mainly detected
through strongly redshifted K$\alpha$ and K$\beta$ absorption lines
from H- and He-like iron. See, e.g., \citet{chartas:02a},
\citet{pounds:03a}, \citet{turner:04a}, or \citet{cappi:09a} for
discussions of individual sources and \citet{Tombesi2010a} for a
recent comprehensive study with \textsl{XMM-Newton}.

Radiation hydrodynamical calculations show that such winds can in
principle be formed as line driven winds from an accretion disc
\citep[][and references therein]{proga:00a,kurosawa:09a}. The spectral
signature imprinted on an X-ray continuum are in rough agreement with
the features seen here \citep[e.g.,][]{schurch:09a}, although further
theoretical work such as proper inclusion of Compton broadening is
clearly needed \citep[e.g.,][]{sim:10a}. Unfortunately, the current
wind models available for X-ray spectral modeling also do not yet
allow us to self-consistently model absorption with abundances
consistent with the significant overabundance in the accreted material
inferred from the X-ray reflection. Despite this problem, however, the
agreement between our simple wind model and the data is remarkable. In
passing, we note that the measured outflow velocity is also consistent
with the unified model for quasars proposed by \citet{Elvis2000a},
where a ``Warm Highly Ionised Medium'' is ejected at speeds of
10000--$60000\,\mathrm{km}\,\mathrm{s}^{-1}$ at an inclination of
around $60^\circ$. Coincidentally, this value is in agreement with our
best-fit inclination angle. Note that in order to explain the apparent
shift of minor spectral features in the 2--5\,keV band, the line of
sight velocity of this wind must have changed between both
observations. As such a wind is expected to be highly structured
\citep{sim:10a}, only very slight changes in the line of sight would
be required to explain the observed change in velocity.

Despite the overall success of the modeling, however, some significant
broad residuals remain (Fig.~\ref{fig:spectra}d). There are several
major issues which could explain these discrepancies that could not be
treated properly in our analysis due to limitations of our best-fit
model (Sect.~\ref{sec:basic-model}). First of all, while measurements
of the reflection show a high Fe abundance in reflection, all other
elements are assumed to be of Solar abundance. From basic arguments it
is clear that a high Fe abundance likely implies that other elements
are over-abundant as well, which could not be taken into account in
our approach due to the limitations of the \texttt{reflionx} model.
Such a restriction might therefore underpredict emission from other
elements, while at the same time over-predict the Fe abundance.
Improving on modeling the ionisation gradient in the accretion disc as
discussed above could also slightly reduce the artificially high Fe
abundance obtained from fitting the data with simple reflection models
\citep{Reynolds1995a}. As already discussed by \citet{Zoghbi2010a},
however, NLS1 galaxies such as 1H0707$-$495 exhibit enhanced star
formation \citep{Sani2010a} and therefore are expected to be Fe
enriched. This assumption is confirmed in near-IR measurements
\citep[see, e.g., ][]{Shemmer2002}. Moreover, in similar sources like
MCG$-$6-30-15 \citep{Miniutti2007amnras}, 1H0419$-$577
\citep{Fabian2005a}, or IRAS13224$-$3809 \citep{Ponti2010a} iron is
also required to be over-abundant.

 Regarding the ionised wind seen in absorption, we measure a lower wind
velocity and $N_\mathrm{H}$ compared to earlier absorption dominated
models \citep{Gallo2004a,done2007a}. This difference is probably due
to the fact that in earlier models the full continuum was seen to be
dominated by the wind, while in the present model the soft-excess is
mainly described by reflection of a highly ionised accretion disc.
This interpretation is in line with our identification of weaker
spectral features as absorption lines from H-like ions (Si, S, Ca, and
possibly Ar), which are found at the correct energies expected from
the inferred wind velocites. As discussed above, since the 7\,keV
feature is non variable, it cannot be due to the wind. This result
significantly simplifies the wind modeling, as in order to explain the
whole 7\,keV feature as a wind a very complex absorber is necessary
\citep[see also][note, however, that the ionised and blue-shifted
absorption is required to describe the narrow absorption feature right
above the drop]{done2007a}.

\emph{Acknowledgements.} This work is based on observations obtained
with \textsl{XMM-Newton}, an ESA science mission with instruments and
contributions directly funded by ESA Member States and NASA. JW and TD
acknowledge partial support from the European Commission under
contract ITN 215212 ``Black Hole Universe'' and TD by a fellowship
from the Elitenetzwerk Bayern. JS acknowledges support from the Grant
Agency of Czech Republic (GACR 202/09/0772). We thank John Davis for
the development of the \texttt{SLxfig} module used to prepare the
figures in this paper, and Alex Markowitz and Moritz B\"ock for their
many useful discussions.
 
\bibliographystyle{mn2e_williams} \bibliography{mnemonic,mn_abbrv,bib}

\appendix
\label{lastpage}

\end{document}